\documentclass[rmp,showpacs,twocolumn,floatfix]{revtex4}

\usepackage{dcolumn,graphicx,amsmath,amssymb,txfonts}

\begin{document}

\title{Local symmetries in complex networks}

\author{Petter Holme}
\affiliation{Department of Computer Science, University of New Mexico,
  Albuquerque, NM 87131, U.S.A.}

\begin{abstract}
  Symmetry---invariance to certain operators---is a fundamental
  concept in many branches of physics. We propose ways to measure
  symmetric properties of vertices, and their surroundings, in
  networks. To be stable to the randomness inherent in many complex
  networks, we consider measures that are continuous rather than
  dichotomous. The main operator we suggest is permutations of the
  paths of a certain length leading out from a vertex. If these paths
  are more similar (in some sense) than expected, the vertex is a
  local center of symmetry in networks. We discuss different precise
  definitions based on this idea and give examples how different
  symmetry coefficients can be applied to protein interaction
  networks.
\end{abstract}


\maketitle

\section{Introduction}

Since the turn of the century, the field of complex networks has been
one of the most active areas of statistical
physics~\cite{ba:rev,mejn:rev,doromen:book,lato:rev}. One of the central
questions is to find quantities for measuring network
structure (how a network differs from a random graph). The basic
assumption is that the network structure is related to the function of
the network. Thus, by measuring network structural quantities, one
can say something both about the forces that created the network, and
about how dynamic systems on the network behave. One important concept in
many areas of physics (particle physics, condensed matter physics and
more~\cite{my:sym}) is symmetry---invariance to particular operators. Our
approach is to presuppose that symmetry can be useful to
study complex networks, then we try to construct a sensible and
general framework for measuring symmetry in networks.

\begin{figure}
  \resizebox*{0.9 \linewidth}{!}{\includegraphics{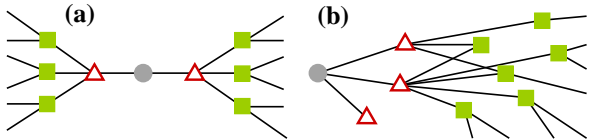}}
  \caption{ An illustration of perfect degree symmetry (a) and perfect
    symmetry of external traits (b). In (a) different symbols
    represent different degrees. In (b) a symbol represents an
    external trait (in this paper we exemplify this by functional
    categories of proteins). (Non-self-intersecting) paths of
    length two will, in both cases, first lead to a triangle,
    then to a square, which means the circle is a center of symmetry.
  }
  \label{fig:ill}
\end{figure}

In Ref.~\cite{my:sym} we define a measure for degree-symmetries in
networks---a degree-symmetry coefficient. This is a local, vertex-specific
measure, i.e.\ it includes only information
from a bounded surrounding of the vertex. The fundamental operator in
this definition of degree-symmetry is permutations of paths of length
$l$ leading out from a vertex $i$. If the degree sequences of paths of length
$l$ from $i$ overlap to a great extent, then we say $i$ is a center of
degree-symmetry. In other words; if, regardless of which path we take
out from $i$, we see the same sequence of degrees, then $i$ is highly degree-symmetric. If one replaces degree, in this
definition, by some other vertex-specific quantity, one gets a
general framework for analyzing local symmetry---instead of degree
symmetries, one can talk about clustering symmetries, betweenness
symmetries or symmetries with respect to any other (network related or
external) vertex specific quantity. (See Fig.~\ref{fig:ill}.) In this
paper we will discuss such extensions of degree-symmetry
coefficient. As one example we study functional symmetries in networks of proteins.

\section{Definition of the measure}

We consider a network modeled by an unweighted and undirected graph of
$N$ vertices, $V$; and $M$ edges, $E$. We assume the graph have no multiple edges or self-edges. Let $X(i)$ be a
vertex trait or structural quantity---for example: degree, betweenness
centrality~\cite{mejn:rev,doromen:book,lato:rev} or a protein function. Consider a vertex
$i$ and the paths of length $l$ leading out from this vertex. These
paths can be thought of as the look of the network from the vantage
point $i$. The cut-off length $l$ reflects that the influence of the
network $i$ on $i$'s function decreases with distance. In
principle one can use any decaying function to lower the weight of
distant vertices. We chose the simplest functional form (at least the
easiest to implement)---a step function weighing vertices at a
distance $l$, or less, from $i$ equal (while yet more distant vertices are not considered at all). In the numerical examples, we will choose the
shortest non-trivial value, $l=2$. The sequences of $X(i)$-values
along these paths are the input to the symmetry measure. We denote such
sequences:
\begin{eqnarray}
  Q^X_l(i)&=&\Big\{[X(v^1_{1,i,l}),\cdots,X(v^l_{1,i,l})],\nonumber\\
&&\vdots\\
 && ,[X(v^1_{p,i,l}),\cdots,X(v^l_{p,i,l})]\Big\},\nonumber
\end{eqnarray}
where $v^j_{m,i,l}$ is the $j$'th vertex along the $m$'th path of
length $l$ leading out from $i$. Then let $F(X,X')$ be a function
measuring the similarity of two $X$-values (for integer valued
$X$-functions, one example of an $F$-function is Kronecker's delta). A
first attempt to construct a symmetry measure is to sum $F(X(i),X(j))$ for
vertex pairs at the same distance from $i$ in $Q^X_l(i)$, i.e.\
\begin{equation}\label{eq:para}
  \frac{\tilde{s}_l(i)}{\Lambda}=\sum_{0\leq n<n'\leq p}\sum_{j=1}^l
    F\big(X(v^j_{n,i,l}), X(v^j_{n',i,l})\big),
\end{equation}
where
\begin{equation}\label{eq:lambda}
\Lambda=(l-1)\:\dbinom{p}{2} .
\end{equation}
This measure has many statistical discrepancies. For example, all
paths that go via a particular neighbor of $i$
contribute to the sum. In practice this means that vertices with a
high degree vertex $\hat{i}$ at a distance close to $l$ will (by virtue of the many paths that
overlap up to $\hat{i}$) trivially
have a high $\tilde{s}_l(i)/\Lambda$. To get around this problem we omit path segments at
indices lower than $\hat{i}$ in $Q^X_l(i)$ (for details, see Ref.~\cite{my:sym}). Let $S_l(i)$
denote the number of such terms (a way to calculate $S_l(i)$ is given
in Ref.~\cite{my:sym}). Then a measure compensating for terms from path
with the same beginnings is given by:
\begin{equation}\label{eq:proto}
  s'_l(i)=\frac{\tilde{s}_l(i)- S_l(i)}{\Lambda- S_l(i)} , \mbox{~~
    provided $\Lambda > S_l(i)$.}
\end{equation}
The degree sequence is often considered an inherent property of the
system. Structure should, in such cases, be defined relative to a
null-model of random graphs conditioned to the same degree
distribution as the network. A measure where zero denotes neutrality
can be constructed as:
\begin{equation}\label{eq:frame}
  s_l(i)=s'_l(i)-\langle s'_l(i)\rangle ,
\end{equation}
where $\langle \:\cdot\:\rangle$ denotes average over an ensemble of random graphs
with the same set of degrees as the original network. A way to sample such null-model graphs is to randomly rewire the
edges of the original network (at every time step keeping
the vertices' degrees are conserved). Note that, for such rewiring procedures, there are many
sample-technical considerations needed to achieve ergodicity and statistical independence. We use the scheme proposed in
Ref.~\cite{roberts:mcmc} and 1000 sample averages. If the $X$-function
only depends on the network, one can recalculate it
for each individual realization of the null-model. If the information
behind $X(i)$ is external, then one has to let the trait be
associated with $i$ throughout the randomization process, or
randomly distribute the traits among the vertices. The former
situation is suitable if the trait has some connection to the degree,
the latter (that we use in this paper) is more appropriate if there are no such connections.

To apply the framework described above one has to specify a function
$X$ mapping $V$ to integer or real numbers. Furthermore one has to
chose an $F$-function indicating if two vertices are considered
similar or not. In this paper we discuss binary valued $F$-functions
($F(X(i),X(j))=1$ if $i$ and $j$ are considered similar,
$F(X(i),X(j))=0$ otherwise), but one can also think of real valued
$F$-functions where a high value means a high similarity between the
two arguments.

\begin{figure*}
  \resizebox*{0.75 \linewidth}{!}{\includegraphics{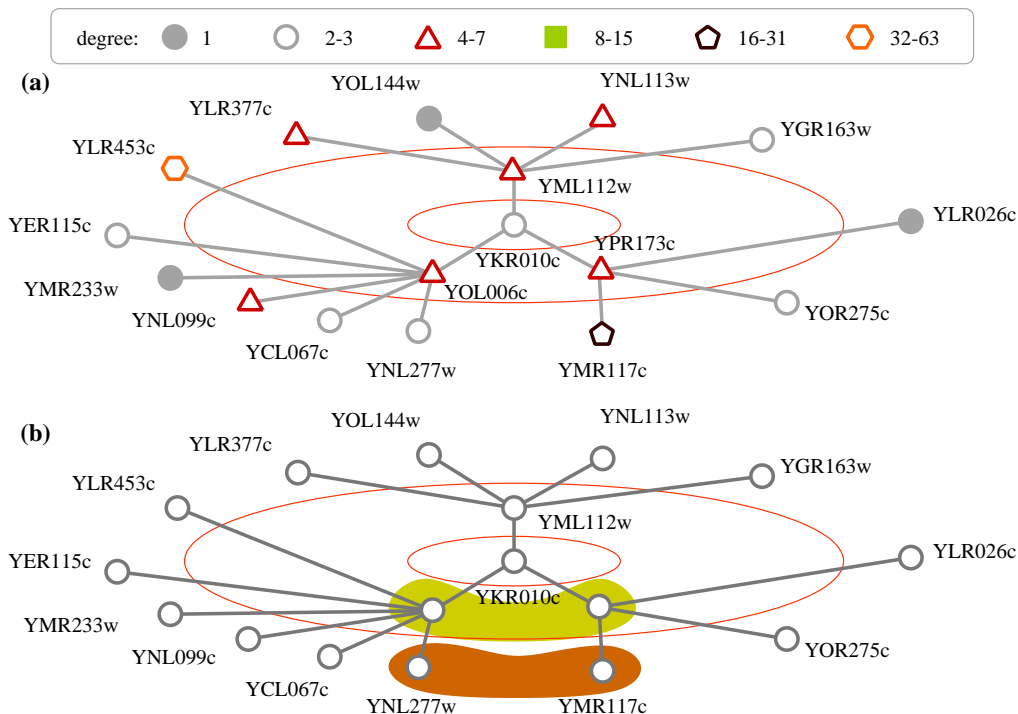}}
  \caption{
  Example from the \textit{S.\ cerevisiae} protein interaction network
  illustrating the symmetries of YKR010c. The concentric
  ellipses mark the first and second neighborhoods. (a) illustrates
  the configuration giving the symmetry coefficient $0.809$. (b)
  illustrates the functional symmetries resulting in a functional symmetry
  coefficient of $0.299$. The vertices connected by a shaded area have the
  identical sets of functions.
  }
  \label{fig:ykr010c}
\end{figure*}

\begin{figure*}
  \resizebox*{0.82 \linewidth}{!}{\includegraphics{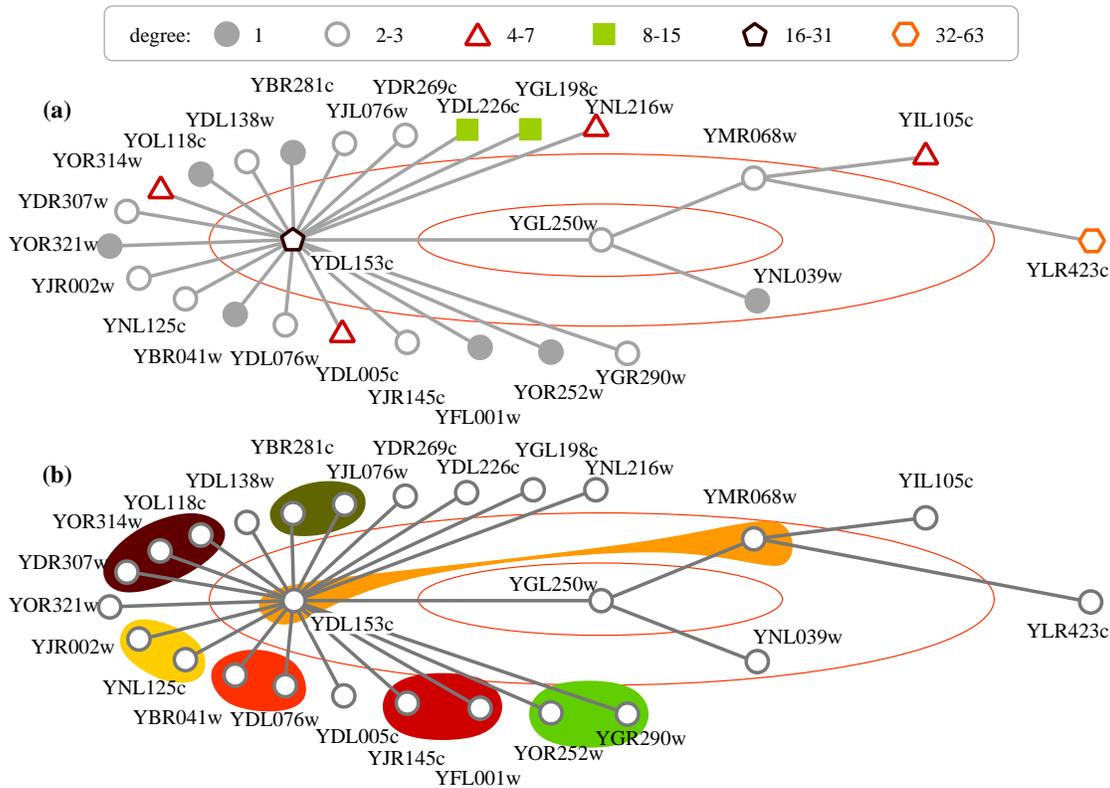}}
  \caption{ The two-neighborhood of YGL250w in the \textit{S.\
      cerevisiae} protein interaction network. The symbols are the
    same as in Fig.~\ref{fig:ykr010c}. (a) shows the degree symmetry
    situation giving the symmetry coefficient $-0.178$. (b) shows the
    functional overlaps in the two-neighborhood of YGL250w giving a
    functional symmetry coefficient of $0.965$.
  }
  \label{fig:ygl250w}
\end{figure*}

\section{Applications to protein interaction networks}

One of the most successful applications of complex network analysis
is studies of large-scale microbiological networks. Such studies can
be performed at different levels of the cellular organization---from
genetic regulation~\cite{david:gene,wagner:robu}, via protein
interactions~\cite{janin:pin,yook:protein}, to biochemical
networks~\cite{wagner:robu,zhao:meta}. We will use protein interaction
networks as our example. In protein
interaction networks the vertices are typically an entire proteome (i.e.\ all
proteins in an organism). The edges represent pairs of
proteins than can bind physically to each other. It is important to
note that at only a small fraction of the protein interactions is in
effect at particular location in a particular cell. The biological
information one can hope to get out from studying the protein
interaction network is thus rather limited.  Dynamic properties of
the cellular activity, i.e.\ the functions of a particular cell, are
beyond the reach of static network theory. The study of the protein
interaction network, in this paper, serves more as an example of
symmetry analyzes, than an advance in proteomics. If
symmetry has some relation to the protein functions, like degree is
correlated with lethality~\cite{jeong:leth}, one can use the symmetry
coefficient for functional classification or prediction.

The particular protein interaction data we use (from the yeast
\textit{S.\ cerevisiae}) was taken from MIPS~\cite{pagel:mips} January
23, 2005 (the same
data set as used in Ref.~\cite{hh:pfp}). The network has $N=4580$ and
$M=7434$. MIPS also provide functional classification of the
proteins~\cite{ruepp:mips}. This is a hierarchical classification
where, for example, the top-level category ``metabolism'' is
subdivided into e.g.\ ``amino acid metabolism,'' and so on. One
protein can be assigned none, one or many functional categories; so,
to make a symmetry measure out of this information, let $X(i)$ be the
set of top-level functions of $i$, and let
\begin{equation}\label{eq:proteinf}
 F(X,X')=\left\{\begin{array}{rl} 1 & \mbox{if $X=X'$}\\ 0 &
     \mbox{otherwise}\end{array}\right . .
\end{equation}
We choose this $F$-function because it is the simplest. For a more
thorough investigation of protein interaction symmetries one might
consider other functions, like the real valued Jaccard-index.

Apart from the functional symmetry coefficient we will also measure the
degree-symmetry coefficient as in Ref.~\cite{my:sym}. In this case $X(i)$
is the degree, or number of neighbors, of $i$. For highly skewed
degree distributions, as protein interaction networks are known to
have~\cite{jeong:leth}, it is appropriate to use:
\begin{equation}
F(k,k') = \left\{\begin{array}{rl} 1 & \mbox{if $\exists i$ such that
      $a^i\leq k,k'<a^{i+1}$}\\ 0 & \mbox{otherwise}\end{array}\right
. 
\end{equation}
We use $a=2$ and $i=0,1,2,3,\cdots$.

In Fig.~\ref{fig:ykr010c}(a) we give an example of a protein with high
degree symmetry, YKR010c. Since its neighbors are all equal (i.e.\ all
pairs of neighbors $(i,i')$ have $F(k_i,k_{i'})=1$) this is not
surprising. Even many second-neighbors are equivalent in this respect
(such as YLR377c, YNL113w and YNL099c). Fig.~\ref{fig:ykr010c}(b)
shows the functional overlap in the same subgraph. Although the
overlapping vertex pairs are rather few, YKR010c has a positive
functional symmetry coefficient (rather weak, however, with a p-value of
around five percent). The main reason for this is that similar
vertices are very rare due to the quite strict definition of
similarity (Eq.~\ref{eq:proteinf}). Fig.~\ref{fig:ygl250w}(a) shows a
protein, YGL250w, with a negative degree-symmetry coefficient. The
visual impression of skewness of YGL250w's two-neighborhood is, we
believe, another aspect of this degree-asymmetry. In contrast, the
functional symmetry coefficient of YGL250w vertex is large. As noted
above, due to the many possible sets of functions (675 in total)
functionally overlapping pairs are quite rare; yet in this example there are seven sets of overlapping pairs, or triplets at the same
distance from YGL250w which explains the high functional symmetry.

\section{Discussion and conclusions}

In this paper we have proposed a general framework for measuring
symmetries of the surrounding of a vertex. The basic idea is that
observational processes often take the form of walks; in other words,
that the symmetry means that the network looks the same along many paths leading out from a vertex. This leads us to the
principle that if the set of paths of a limited length $l$ out from
a vertex $i$ is invariant to permutations, then $i$ is a local center
of symmetry. We exemplify this framework, and the derived symmetry
coefficient, by studying the protein interaction network of \textit{S.\
  cerevisiae}. For this network databases catalog traits of the
vertices, which allow two fundamentally different symmetries to be
measured: the degree symmetry (where the similarity is
related to the network structure) and functional symmetry (where the
similarity stems from external information). These two coefficients
are exemplified by two proteins in very different symmetry
configurations (one with high degree symmetry and weakly positive
functional symmetry, another with degree asymmetry and very
high functional symmetry). We do not attempt to deduce the biological
meaning of the symmetry coefficients. But we can conceive that
symmetry and biological function are related from the presence of
``network-motifs''~\cite{alon} in biological networks. Network
motifs are small, statistically overrepresented subgraphs with,
presumably, specific functions. If one vertex controls, or is
controlled by, several such motifs, then it would have high (degree,
functional or other) symmetry coefficient. To conclude, we believe
symmetries can be a useful concept for analyzing complex
networks. There are, furthermore, many ways to extend this work to
other measures and applications.

\acknowledgements{
  PH acknowledges financial support from the Wenner-Gren foundations,
  the National Science Foundation (grant CCR--0331580), and the Santa
  Fe Institute.
}

\end{document}